\documentclass[a4paper,10.5pt,twocolumn,oneside]{article}

\usepackage{graphicx}
\usepackage{color}
\usepackage[a4paper, left=1.4cm, right=1cm, top=1.7cm, bottom=1cm]{geometry}
\usepackage[super,comma,sort&compress,square]{natbib}
\usepackage{amsmath}
\usepackage{float}
\usepackage{latexsym}
\usepackage{amssymb}

\usepackage{fancyhdr}
\usepackage[utf8x]{inputenc}
\usepackage[T1]{fontenc}
\begin{document}
\pagestyle{fancy}
\def\headrulewidth{0.5pt}
\def\footrulewidth{0pt}
\lhead{Journal of Alloys and Compounds 646 (2015), 773 -- 779} 
\chead{}
\rhead{DOI: 10.1016/j.jallcom.2015.05.190}

\lfoot{} 
\cfoot{}
\rfoot{}

\twocolumn[
  \begin{@twocolumnfalse}
  {\huge \bf Structural ordering of laser-processed FePdCu thin \mbox{alloy} films}

  \hspace{1.1cm}
  \parbox{.87\textwidth}{
    \vspace{4ex}
%    \begin{center}
    \Large \textsf{Marcin Perzanowski, Michal Krupinski, Arkadiusz Zarzycki, Yevhen Zabila, Marta Marszalek}
    \vspace{1ex} \\
    \normalsize The Henryk Niewodniczanski Institute of Nuclear Physics, Polish Academy of Sciences, Radzikowskiego 152, 31-342 Krakow, Poland
%    \end{center}
    \vspace{1ex} \\
    \normalsize \text{email: Marcin.Perzanowski@ifj.edu.pl}

    \vspace{2ex} 
    \noindent
     \textbf{Abstract}: The Cu/Fe/Pd multilayers were transformed into L10-ordered FePdCu alloy by pulsed laser annealing. 
     The initial multilayers were irradiated with 1, 10, 100, and 1000 laser pulses with duration time of 10 ns and energy density of 235 mJ/cm$^{2}$. The gradual change of the number of laser pulses allowed to investigate the structural and magnetic properties at early stages of the transformation and L1$_{0}$-ordering processes. The measurements were carried out using X-Ray Diffraction, SQUID magnetometry, and Magnetic Force Microscopy. We found that L1$_{0}$ FePdCu (111)-oriented nanograins are formed by ordering of the coherent domains present in the as-deposited multilayer. The irradiation does not change the vertical size of the (111) crystallites. The L1$_{0}$ (002)-oriented grains appear at the later stages of the transformation and their size increases with the number of applied laser pulses. Additionally, the laser annealing induces the magnetic ordering of the irradiated material, which was observed as an increase of the saturation magnetisation and the Curie temperature with the rising number of pulses. We also observed, that irradiation with 1000 pulses leads to the loss of order, which is reflected in the drop of the Curie temperature.

     \vspace{2ex}
     DOI: 10.1016/j.jallcom.2015.05.190 
     
     \vspace{2ex}
     
    \vspace{3ex}
  }
  \end{@twocolumnfalse}
]

\section{Introduction}

\vspace{-0.4cm}

The development of new materials for the magnetic storage industry requires investigation of the new methods of their fabrication. 
Magnetic thin alloy films with L1$_{0}$ crystal structure such as FePd, FePt and CoPt, which exhibit a large uniaxial magnetic
anisotropy of a~few MJ/m$^{3}$,\cite{1,2,3,4} are considered as potential materials for the bit pattern high-density recording media with information density over 1~Tbit/in$^{2}$. 
Application of these materials in data storage devices with perpendicular recording requires the fabrication of (001)-textured film.\cite{5} 
In order to design a~material with desired properties it is essential to investigate the mechanisms leading to the formation of the L1$_{0}$-ordered alloy. 
In the following paper we focus on the FePd alloy: it is less expensive than the other two Pt-based materials; it has a lower order-disorder transition temperature\cite{6} while having similar saturation magnetisation; and it is considered a~model system for an exchange coupled nanomagnets.\cite{4}

One possible way to obtain an L1$_{0}$-ordered FePd thin film is an epitaxial growth.\cite{7,8,9,10} 
However, this method is relatively complicated and therefore it is hard to apply it for mass production. 
Another way to fabricate a film is the co-deposition of the constituent materials,\cite{11,12} or the deposition of the multilayer\cite{13} and its transformation into L1$_{0}$-ordered alloy. 
The latter can be done by the conventional long annealing,\cite{14} rapid thermal annealing,\cite{15} or by the ion beam irradiation.\cite{7} 
In this paper we present a new approach for fabrication of the L1$_{0}$-ordered FePd thin films, which is based on the pulsed laser annealing. 
Among the magnetic materials with L1$_{0}$ structure only for FePt thin alloy films results were previously obtained with this method by application of microsecond or sub-microsecond laser pulses showing that it leads to the formation of an alloy with the desired crystal structure.\cite{16,17}

The application of the laser annealing gives high heating and cooling rates of the irradiated material, which cannot be obtained
with other methods. 
The increase of a~number of laser pulses allows the investigation of the material properties at the consecutive early
stages of the transformation between a multilayer and an alloy film. 
Information gained during these studies is essential for understanding of the processes leading to the creation of the ordered L1$_{0}$ alloys with particular crystallographic texture, which is of great interest from the point of view of materials engineering. 
Moreover, such results can be exploited in experiments on Direct Laser Interference Patterning of metallic thin films.\cite{18}

The FePd thin films investigated here have an addition of 10 at\% of copper, since it was found that such addition lowers the ordering temperature\cite{19} and facilitates the fabrication of an L1$_{0}$-ordered alloy.\cite{20,21,22}

\vspace{-0.6cm}
\section{Experimental details}

\vspace{-0.3cm}
The [Cu(0,2 nm)/Fe(0,9 nm)/Pd(1,1 nm)]$_{10}$ multilayers were deposited by thermal evaporation on the Si(100) substrates with a~native SiO$_{x}$ oxide layer. 
The thicknesses of Fe and Pd layers were chosen in order to obtain equiatomic FePd alloy, and the Cu thickness corresponded to atomic amount of 10\%. 
Before the deposition the Si substrates were ultrasonically cleaned in acetone and ethanol, and rinsed in de-ionised water. The pressure in the evaporation chamber during the deposition was of the order of 10$^{-7}$~Pa. 
The constituent layers were deposited sequentially, with the evaporation rates of 0,5~nm/min for Fe and Pd, and 0,2~nm/min for
Cu. 
The thickness of layers was controlled in-situ by quartz crystal microbalance. The samples were 5 mm by 5 mm in size in order to
fit with the diameter of a~laser spot. 
After the deposition the  chemical composition of the evaporated multilayers was checked by Rutherford Backscattering Spectrometry, confirming the assumed stoichiometry with an average accuracy of $\pm4$~at.\%. 
The thickness and roughness of the deposited layers were verified by X-Ray Reflectometry (XRR). 
The measurements confirmed the nominal thicknesses of the layers with average accuracy of 5\%. 
XRR experiments also showed that the roughness of a~single layer was approximately 0,6~nm. 
This result indicates the presence of a~significant intermixing between layers during the deposition process and suggests that deposited volume of Cu is not a~continuous layer and all Cu atoms are mixed with surrounding Fe and Pd.

In order to form FePdCu thin alloy films the as-deposited Cu/Fe/Pd multilayers were annealed using the pulsed laser irradiation. 
The irradiation was carried out using the multimode Quantel YG980 Nd:YAG laser operated at first harmonic with a~wavelength of
1064 nm. 
The duration of a~single pulse was 10 nm with a repetition frequency of 10 Hz. 
The spatial divergence of an unfocused beam was 0,07~rad, and the beam spot diameter, measured at half of the maximum intensity, was 4,5~mm. 
The energy density of a~laser beam was set by the Q-switch device and was controlled by Coherent FieldMax II energy meter. 
The energy distribution in the beam spot was not perfectly homogeneous; however, the irradiation with an increasing number of pulses should result in an uniform areal distribution of the deposited energy. 
Initial multilayers were irradiated with 1, 10, 100, and 1000 laser pulses. 
To prevent oxidation the annealing process was carried out in atmosphere of a~flowing nitrogen. 
The energy density was 235 mJ/cm$^{2}$ per single pulse. 
The $\Theta$/2$\Theta$ X-Ray Diffraction (XRD) patterns were collected using X'PertPro PANalitycal diffractometer, equipped with X-ray source with Cu anode ($K_{\alpha 1}$ line, wavelength $\lambda$=0,154~nm) operated at 40~kV and 30~mA. 
Experimental conditions were the same as described in Ref.~[23]. 

The measurements of magnetisation as a~function of the external magnetic field (hysteresis loops) as well as the temperature measurements of the magnetisation (field cooling, FC) were carried out using Quantum Design MPMS XL SQUID device. 
The hysteresis loops were collected at 300~K and FC curves were measured in temperature range from 5~K up to 300~K with step of
3~K. 
For hysteresis loops the external magnetic field step was 5~Oe, 100~Oe, 500~kOe, and 5~kOe and was dependent on the field range
and on the measurement geometry. 
The FC curves were acquired for in-plane geometry with the external magnetic field of 100~Oe.
Hysteresis loops were measured for in-plane and out-of-plane geometries, with the external magnetic field directed at angles between 0$^{\circ}$ and 90$^{\circ}$ with respect to the sample plane.

The atomic and magnetic force microscopy (AFM/MFM) images were taken using Digital Instruments NanoScope device. 
The AFM measurements were done in the contact mode, while the MFM scans were collected in the non-contact phase contrast mode. 
In the MFM the distance between the probing tip and the sample surface was approximately 200~nm, the frequency of unloaded cantilever free oscillations was 73~kHz. 
The probing tip was magnetised vertically with respect to the sample plane. 
Both AFM and MFM images were collected sequentially for the same area on the sample surface. 
The probing frequency was 0,6~Hz.

\vspace{-0.6cm}
\section{Results and discussion}

\vspace{-0.3cm}
The energy of the laser pulses was transferred by electronic excitations to the crystal lattice, which led to the rise of the sample
temperature and enabled the rapid diffusion processes within the irradiated material. T
he applied laser energy density corresponded to the maximal temperature of the sample surface of approximately 500$^{\circ}$C. 
This temperature is necessary for transformation of the Cu/Fe/Pd multilayer into an ordered FePdCu alloy with L1$_{0}$ structure.\cite{14}
The temperature during irradiation was estimated using theory of the pulsed laser annealing and a~heat transfer equation.\cite{24}
The absorption coefficient for wavelength of 1064~nm, essential for the calculations, was 28\% and was determined by optical measurements. 
According to the applied computational model of the annealing process the temperature of the irradiated sample rose from the room temperature up to the maximum value in 10~ns and dropped down to the room temperature before the next pulse was switched on. 
It is worth mentioning that the model applied here does not take into account the temperature dependencies of the material properties, such as the radiation absorption, density, specific heat and thermal conductivity, and assumes that between
the subsequent laser pulses the values of these parameters do not change.

\subsection{Structural ordering}

\begin{figure}[!h]
\centering
\includegraphics[width=0.3\textwidth]{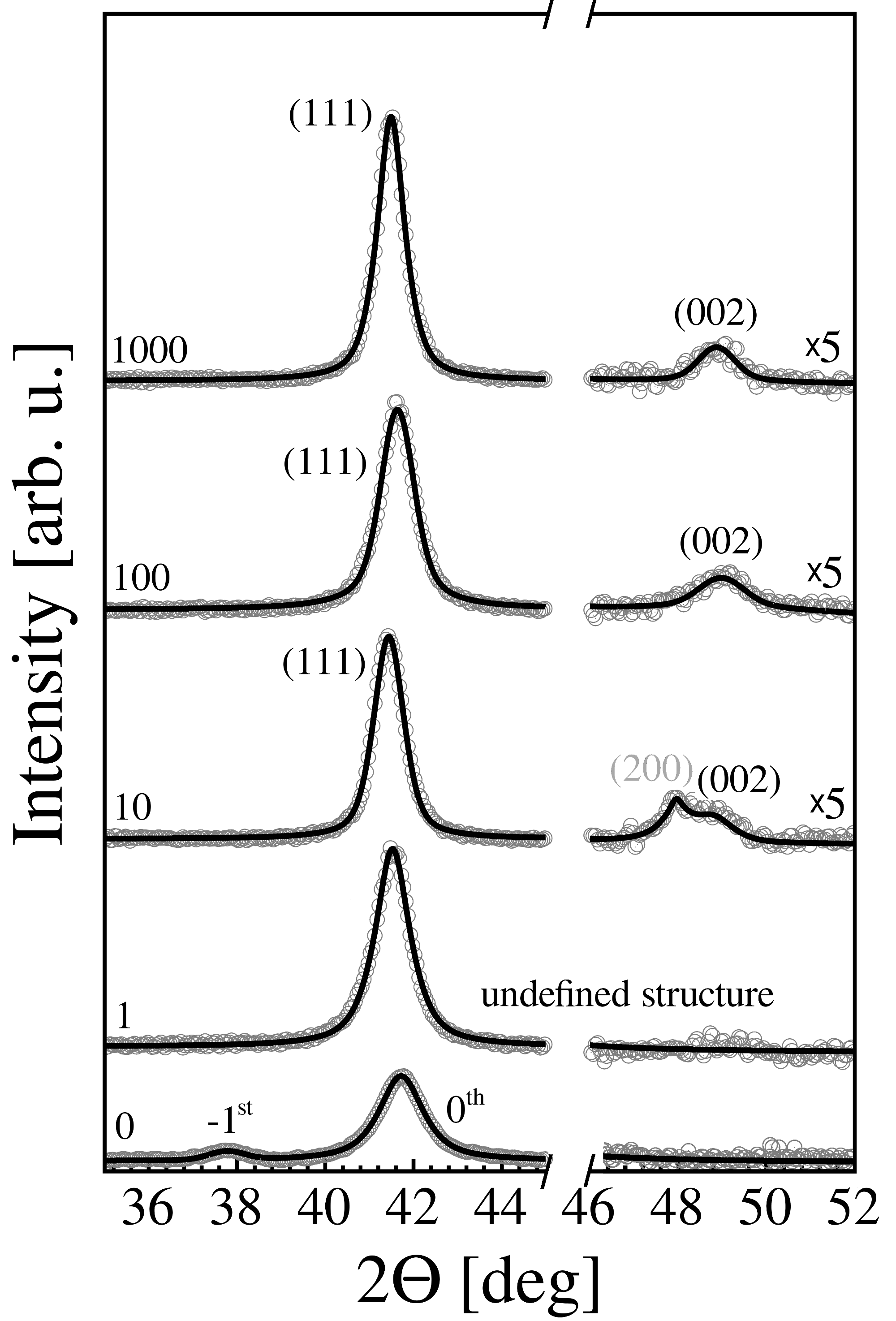}
\caption{XRD patterns of the multilayer and samples irradiated with 1, 10, 100, and 1000 laser pulses. The open circles are the measurement data, the solid lines are fits of the pseudo-Voigt functions. The (111), (002), and (200) reflections from the L1$_{0}$-ordered crystallites are indicated in the figure. Patterns were shifted vertically for clarity.}
\label{fig1}
\end{figure}

The XRD patterns of the as-deposited multilayer and irradiated samples are shown in Fig.~\ref{fig1}. 
For the as-deposited multilayer two reflections at angles $2\Theta=37,6^{\circ}$ and 41,7$^{\circ}$ were observed, and were
identified as -1st and 0th order peaks related to the periodically layered structure of the as-deposited sample.\cite{25,26} 
The angular distance between these two peaks corresponds to the thickness of  the Cu/Fe/Pd trilayer and is $2,24 \pm 0,12$~nm, which is in agreement with the assumed multilayer structure. 
The lack of higher-order satellite reflections was due to the significant layer roughness.
This resulted in the decrease of the peak intensity. 
For the sample irradiated with single pulse, only the reflection at 41,7 was observed. 
The lack of -1st satellite peak is an evidence for the disappearance of the periodic multilayer structure. 
However, the observation of only one reflection makes the precise determination of the crystal structure impossible, since the angular position of this reflection matches both (111) reflection from an L1$_{0}$-ordered alloy and from a~disordered fcc FePd system.\cite{27}

After irradiation with larger number of laser pulses two reflections were observed for $2\Theta$ of 41,7$^{\circ}$ and 48,9$^{\circ}$. 
These peaks were identified as (111) and (002) reflections originating from the L1$_{0}$-ordered crystallites. 
For sample irradiated with 10 pulses the additional reflection was recorded at 47,5$^{\circ}$, and it was identified as the L1$_{0}$ (200) peak. 
The lattice parameters of the irradiated material were calculated using the Bragg equation and the angular positions of the reflections, and they had the values of $a=0,380 \pm 0,001$~nm and $c=0,371 ± 0,001$~nm, giving the lattice distortion $c/a=0,98 \pm 0,01$. 
Both lattice parameters and distortion are close to the values for bulk FePd alloy\cite{27} despite the presence of Cu, although it is known that the Cu addition changes the parameters and increases the lattice distortion.\cite{20,21} 
In this case the observed effect may indicate a~low chemical order in the irradiated samples. 
The appearance of the reflection at 41,7$^{\circ}$ in each irradiated sample means that the structural domains with interplanar distance of about 0,217~nm are present in all cases. 
The effect of the pulsed annealing was the gradual diffusion of the elements leading to the transformation of the as-deposited multilayer into the a thermodynamically stable grained alloy phase with an L1$_{0}$-ordered crystal structure.

\begin{figure}[!h]
\centering
\includegraphics[width=0.25\textwidth]{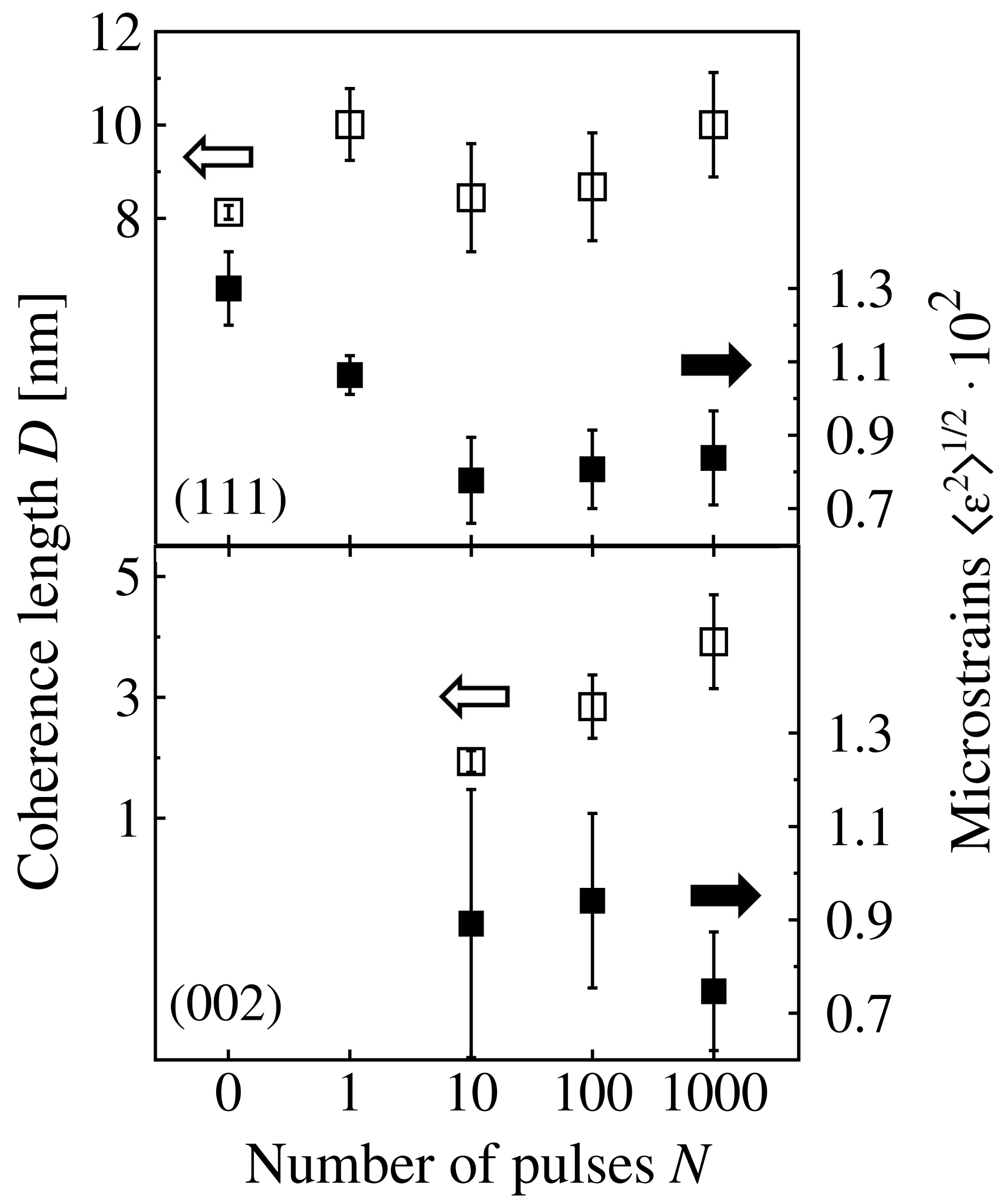}
\caption{Grain sizes $D$ and microstrains $\sqrt{\left< \varepsilon^{2} \right>}$ after different number of laser pulses for (111) and (002) L1$_{0}$-ordered crystallites.}
\label{fig2}
\end{figure}

The observed XRD reflections were fitted with pseudo-Voigt functions. Applying the Scherrer equation
\begin{equation}
 D = \frac{K_{\mathrm{S}} \lambda}{FWHM_{\mathrm{C}} \cos\Theta_{0}} \ ,
\end{equation}
and using the full width at half maximum of the Cauchy component ($FWHM_{\mathrm{C}}$) of the pseudo-Voigt function, the information about coherence length $D$ (grain size along the normal to the sample plane) was obtained. 
The value of the Scherrer constant $K_{\mathrm{S}}$ was 0,95.\cite{28} The wavelength $\lambda$ was 0,154~nm, and the $2\Theta_{0}$ was the angular position of the reflection maximum. 
The microstrains $\sqrt{\left< \varepsilon^{2} \right>}$, being a~root-mean-square of the interplanar distance distribution around
the average value, were calculated using the equation
\begin{equation}
 \sqrt{\left< \varepsilon^{2} \right>} = \frac{w_{\mathrm{G}}}{\sqrt{8 \ln 2} \tan\Theta_{0}} \ ,
\end{equation}
where $2w_{\mathrm{G}}$ is the full width at half maximum of the Gauss component of the pseudo-Voigt function. 
The results of the calculations are shown in Fig.~\ref{fig2}.

The vertical size $D$ of the structural domains with interplanar spacings of approximately 0,217 nm does not change significantly
with an increasing number of laser pulses, and has an average value of approximately 9~nm. 
On the other hand the microstrain $\sqrt{\left< \varepsilon^{2} \right>}$ decreases from $0,013 \pm 0,001$ for the as-deposited multilayer to $0,007 \pm 0,001$ after application of at least 10 pulses. 
This means that the laser annealing results in the strain release for (111) grains, which in turn leads to the structural ordering of the material. 
The (111)-oriented L1$_{0}$-ordered nanocrystallites with narrow height distribution originate from structural superlattice domains present in the as-deposited multilayer. 
For the L1$_{0}$-ordered FePdCu structure the (111) crystallographic plane has the lowest surface energy.
Since the system naturally tends to decrease its total energy, the appearance and ordering of the crystallites with such planes
aligned parallel to the sample surface is favourable. 
After application of more than 10 laser pulses there was no further change of height of the (111)-oriented grains.

The (002)-oriented L1$_{0}$-ordered crystallites appeared after irradiation with at least 10 laser pulses. 
The height of these crystallites increased monotonically with a~number of laser pulses, while no clear relation between number of pulses and microstrains was observed. 
In our previous study\cite{23} we found that (002) crystallites are elongated in the direction parallel to the sample
surface. 
Thus, it can be figured out that these crystallites appeared as the result of the heterogeneous nucleation at the interfaces of
irradiated multilayer. 
This indicates that in case of the (002) crystallites the laser annealing leads to their creation and growth; however, it does not remove the microstrain, which is demonstrated by the lack of relation between a~number of laser pulses and microstrains. 
The appearance of these grains can be attributed to the rapid temperature changes during heating and cooling of the sample, which could promote the occurrence of the nonequilibrium thermodynamic conditions favouring the mechanical stress and formation of (002)-oriented instead of (111)-oriented crystallites.

\vspace{-0.4cm}
\subsection{Magnetic properties}

\begin{figure*}[!t]
\centering
\includegraphics[width=0.5\textwidth]{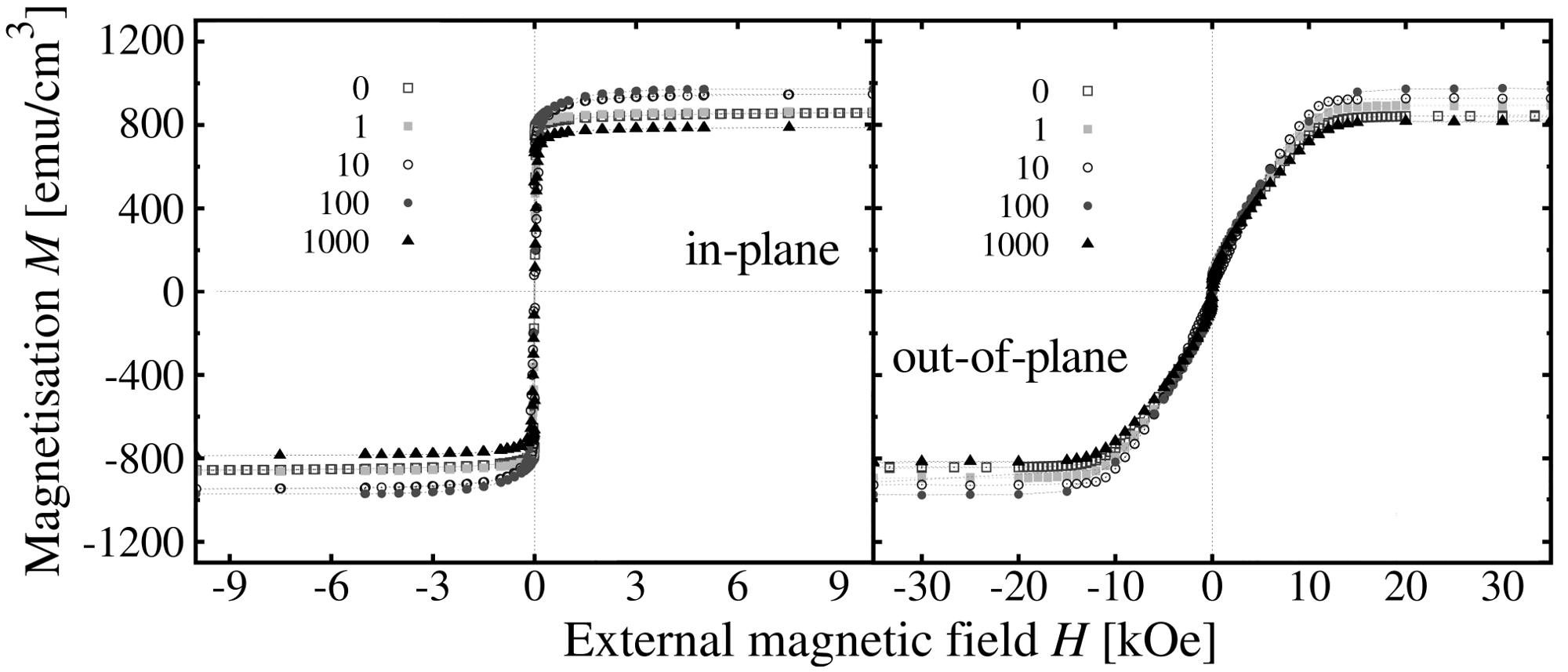}
\caption{The magnetic hysteresis loops $M(H)$ measured for in-plane and out-of-plane geometries of as-deposited and laser-irradiated samples.}
\label{fig3}
\end{figure*}

\vspace{-0.2cm}
Following the laser annealing a~set of magnetic measurements was carried out to obtain information about the change of magnetic
properties. 
The hysteresis loops obtained for out-of-plane and in-plane geometries of the as-deposited multilayer and laser-annealed samples are demonstrated in Fig.~\ref{fig3}. 
For the in-plane geometry the saturation field had value of approximately 3~kOe, while for out-of-plane configuration the value of about 20~kOe was recorded. 
This suggests, that the easy axis of magnetisation is parallel to the sample plane.

\begin{figure}[!h]
\centering
\includegraphics[width=0.35\textwidth]{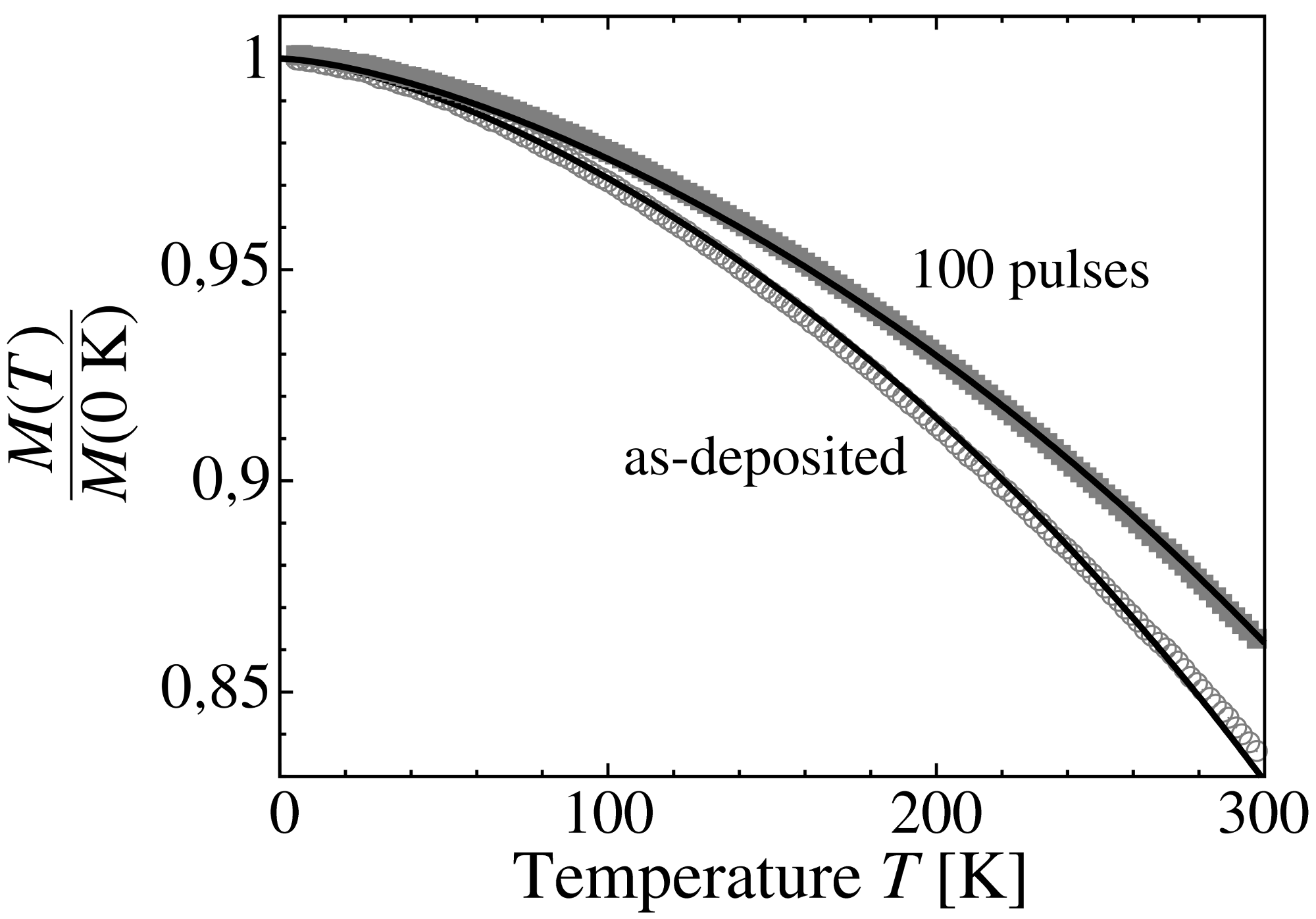}
\caption{An example of the field-cooling measurements for as-deposited multilayer and sample irradiated with 100 pulses, carried out for in-plane geometry with external magnetic field of 100~Oe. Points are data, black solid lines are the fits.}
\label{fig4}
\end{figure}

\begin{figure}[!t]
\centering
\includegraphics[width=0.35\textwidth]{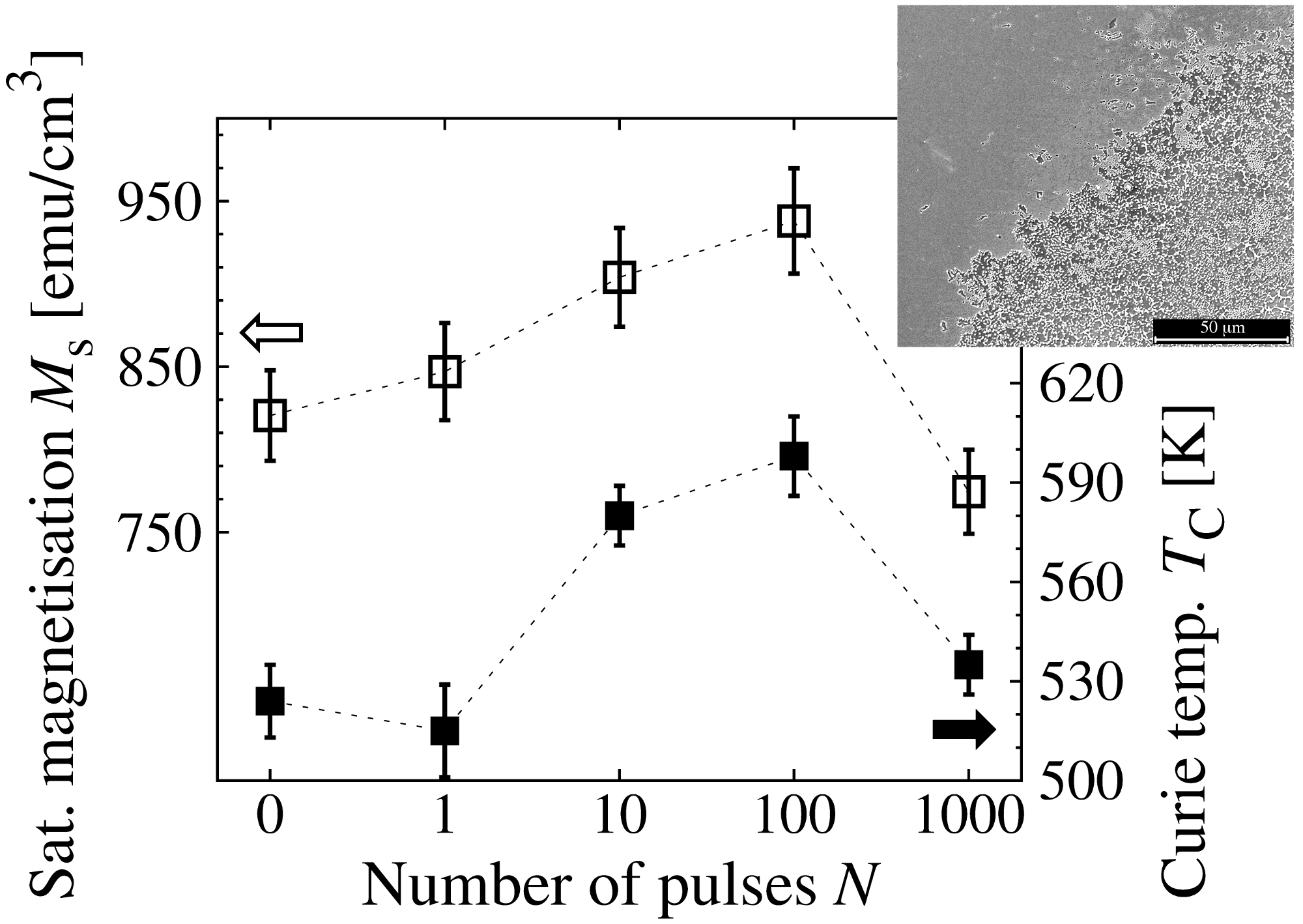}
\caption{The saturation magnetisation $M_{\mathrm{s}}$ and Curie temperature $T_{\mathrm{C}}$ as a~function of a~number of laser pulses. The inset shows the SEM image of the sample surface after 1000 pulses.\cite{23}}
\label{fig5}
\end{figure}

The measurements of the hysteresis loops were supported with the field-cooling experiments, which are shown in Fig.~\ref{fig4}. 
The values of the Curie temperature $T_{\mathrm{C}}$ were obtained from the fit of the field-cooling curves with the function:\cite{29,30}
\begin{equation}
 \frac{M(T)}{M(T=0 \ \mathrm{K})} = \left[  1 - \left( \frac{T}{T_{\mathrm{C}}} \right)^{\frac{3}{2}} \right]^{\beta} \ ,
\end{equation}
where $\beta$ is the critical exponent for ferro-para magnetic phase transition. 
In this case $\beta$ had a~value of $0,33 \pm 0,01$, typical for the ferromagnetic materials and similar to the value obtained for the L1$_{0}$-ordered FePt thin film.\cite{31} 
The values of the saturation magnetisation $M_{\mathrm{s}}$, obtained from hysteresis loops (Fig.~\ref{fig3}), and the Curie temperatures $T_{\mathrm{C}}$, are presented in Fig.~\ref{fig5} as a~function of a~number of laser pulses.
%%%%

It has been reported before that for the well-ordered bulk FePd alloy with L1$_{0}$ structure the magnetic moment of Fe has a~value of 2,85~$\mu_{\mathrm{B}}$/atom, and the Pd atoms have magnetic moment of 0,35~$\mu_{\mathrm{B}}$/atom.\cite{32} 
In our case the magnetic moment on Fe atoms for the as-deposited multilayer, calculated from saturation magnetisation, has a~value of $2,5 \pm 0,1$~$\mu_{\mathrm{B}}$/atom. 
This value is larger than for bulk bcc Fe (2,2~$\mu_{\mathrm{B}}$/atom) and lower than for Fe atoms in an L1$_{0}$-ordered FePd
alloy. 
This indicates that a~fraction of iron atoms created a~disordered FePd phase, which is confirmed by the XRD measurements. 
For the irradiated samples the magnetic moment per iron atom rose up to $2,9 \pm 0,1$~$\mu_{\mathrm{B}}$/atom after 100 pulses. 
The observed increase of the magnetic moment per atom for an increasing number of laser pulses is the evidence for intermixing of the initial layers, and crystallisation of the ordered FePdCu L1$_{0}$ nanograins.

The decrease of saturation magnetisation recorded after 1000 laser pulses was related to the partial ablation of the irradiated
material (see inset in Fig.~\ref{fig5}), which diminished a~number of atoms in the sample. 
A~more detailed study of the laser annealing influence on morphology of the FePdCu thin films was already reported in Ref.~[23]. 
The calculation of saturation magnetisation $M_{\mathrm{s}}$ for the studied samples was based on an assumption that the film is flat and continuous. 
Since there are significant changes in the film morphology, the measured value of the magnetisation was assigned to the larger volume of the film than the measured amount of the material. 
This effect is likely responsible for the lowering of saturation magnetisation after 1000 pulses, although it cannot be excluded that the loss of ordering induced by the laser annealing is also the reason for the lower $M_{\mathrm{s}}$ in this case.

The values of the Curie temperature $T_{\mathrm{C}}$ demonstrated in Fig.~\ref{fig5} are smaller than $T_{\mathrm{C}}$ for an L1$_{0}$-ordered FePd alloy which range from 693~K to 763~K.\cite{33} 
One reason for this can be a~partial ordering of the alloy only, and the presence of the paramagnetic Cu. 
It was found that an increase of a number of laser pulses leads to an increase of the~$T_{\mathrm{C}}$. 
This can be connected to the gradual structural ordering of the irradiated FePdCu alloy. 
The drop of the Curie temperature after 1000 pulses results, as in the case discussed, in the loss of an order of the material, associated with the annealing temperature higher than calculated (see Ref.~[23]) and local ablation of material as seen in inset of Fig.~\ref{fig5}.

The coercivity field~$H_{\mathrm{c}}$ observed in hysteresis loops demonstrated in Fig.~\ref{fig3} shows that for the non-epitaxial [Cu/Fe/Pd]$_{10}$ multilayer, where ferromagnetic Fe layers were decoupled from each other by the non-ferromagnetic Pd/Cu layers with comparable thickness, the coercivity was 13~Oe for in-plane and 60~Oe for out-of-plane. 
It can be expected that after laser treatment inducing the ordering, there will be a~significant increase of~$H_{\mathrm{c}}$. 
However, laser annealing led only to a~slight change of the coercivity resulting in the coercivity of 15 -- 45~Oe for in-plane geometry, and 60 -- 110~Oe for out-of-plane, which was about two orders of magnitude lower than for bulk FePd alloy.\cite{22} 
Since the ordering induced by irradiation may result in the increase of the exchange interaction between magnetic atoms within structural domains, the low value of coercivity recorded after irradiation can be related to the weak dipole magnetic interaction between the newly created L1$_{0}$-ordered grains. 
Therefore, it can be assumed that the L1$_{0}$-ordered nanocrystallites are separated from each other, with disordered material filling the volume between them, which does not provide the significant change in coercivity.

\subsection{Magnetic domains and the mechanism of magnetisation reorientation}

\begin{figure}[!t]
\centering
\includegraphics[width=0.35\textwidth]{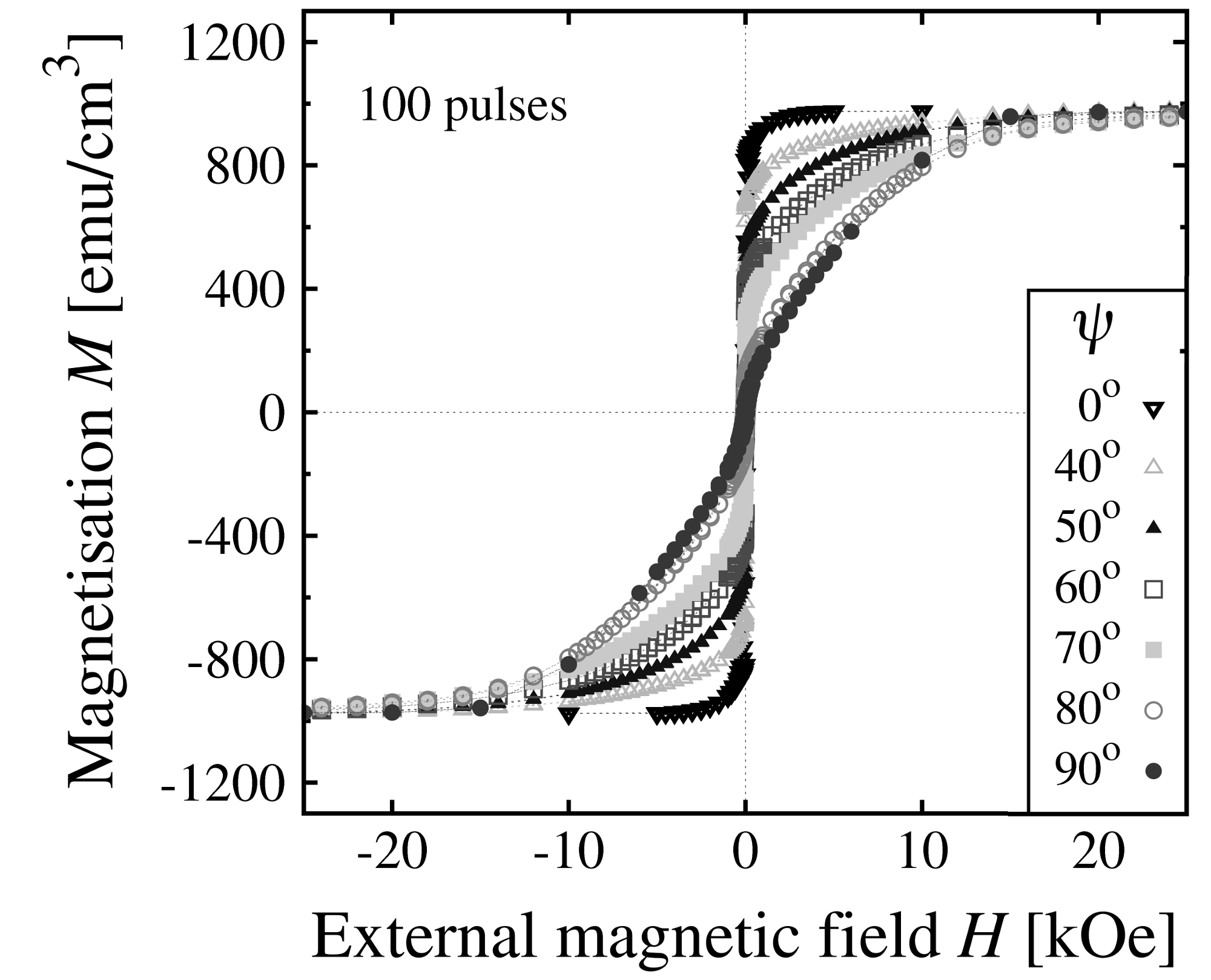}
\caption{Hysteresis loops measured for different angle $\psi$ between direction of the external magnetic field and the direction parallel to the sample plane obtained for the sample irradiated with 100 laser pulses.}
\label{fig6}
\end{figure}

The direction of the easy axis of magnetisation and the mechanism of the magnetisation reversal were measured by the angle-
dependent hysteresis loops for different angles $\psi$ between the direction of the external magnetic field and the sample plane ($\psi = 0^{\circ}$ for in-plane geometry and $\psi = 90 ^{\circ}$ for out-of-plane). 
The results of the measurements are presented in Fig.~\ref{fig6}. 
For the discussion the sample irradiated with 100 pulses was chosen, since it exhibited the largest structural order after laser treatment.

\begin{figure*}[!t]
\centering
\includegraphics[width=0.7\textwidth]{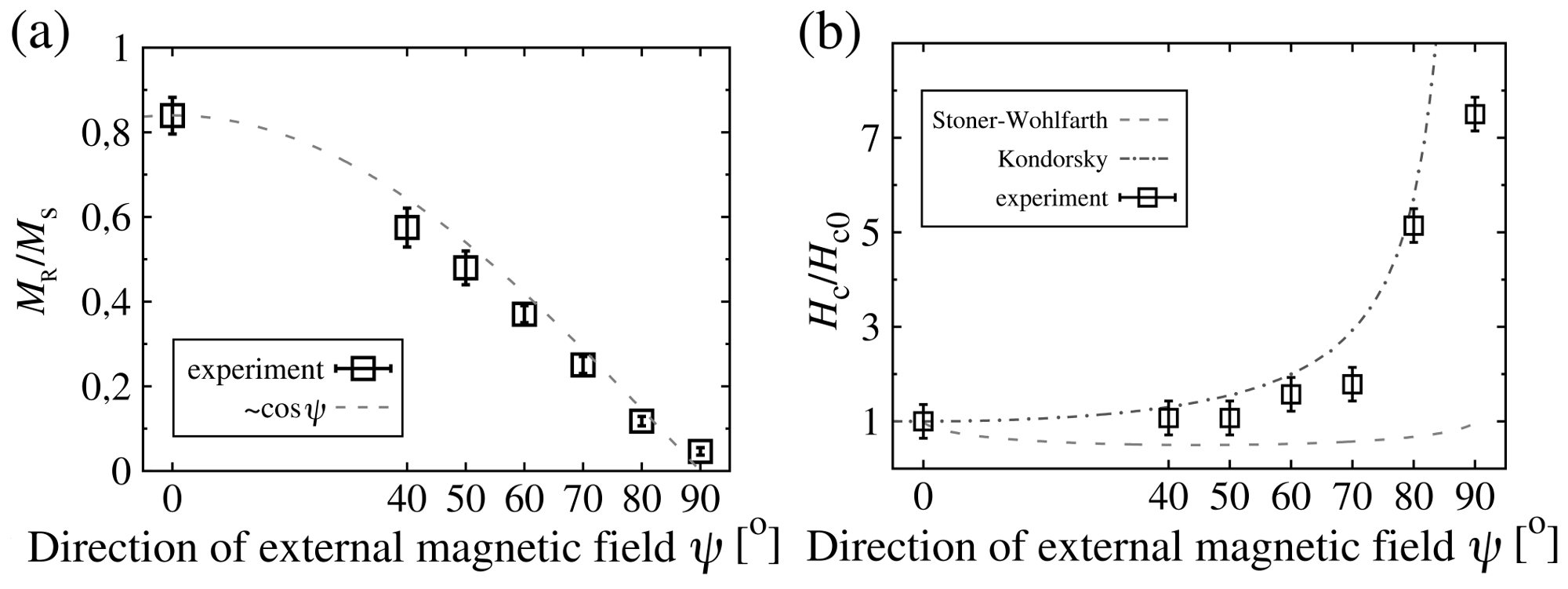}
\caption{(a) The relation between magnetic remanence $M_{\mathrm{R}}$ to saturation magnetisation $M_{\mathrm{s}}$ ratio and angle $\psi$, obtained from the hysteresis loops presented in Fig.~\ref{fig6}. (b) The measured angle-dependent values of the coercivity (open squares), and the theoretical Kondorsky (dot-dashed line)\cite{34,35} and Stoner-Wohlfarth (dashed line)\cite{36,37} relations.}
\label{fig7}
\end{figure*}

The relation between the ratio of magnetic remanence $M_{\mathrm{R}}$ to saturation magnetisation $M_{\mathrm{s}}$ and angle $\psi$ is shown in Fig.~\ref{fig7}a. 
\begin{figure*}[!t]
\centering
\includegraphics[width=0.7\textwidth]{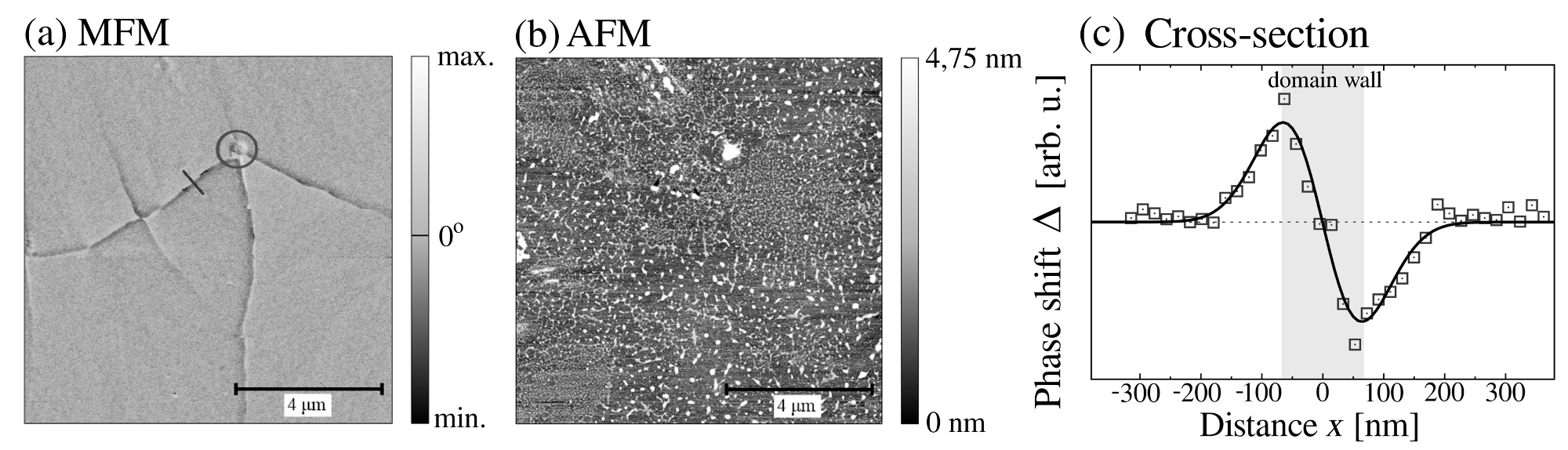}
\caption{Magnetic Force Microscopy (a) and Atomic Force Microscopy (b) images obtained for the same area of the sample irradiated with 100 laser pulses. The circles represent domain wall pinning site. An exemplary cross-section of the magnetic domain wall visible in MFM image, indicated with the line in (a), is presented in figure (c).}
\label{fig8}
\end{figure*}
The change of the $M_{\mathrm{R}}/M_{\mathrm{s}}$ ratio with increasing angle $\psi$ is proportional to cos$\psi$.\cite{38} 
This fact, together with a~monotonic increase of the saturation field as a~function of an angle $\psi$ from approximately 2~kOe to 25~kOe (Fig.~\ref{fig6}), is a~direct evidence that the easy axis of magnetisation is in-plane of the sample. 
The change of the coercivity field Hc with angle $\psi$ is presented in Fig.~\ref{fig7}b. 
The angle-dependent values of the coercivity were normalised to the coercivity value $H_{\mathrm{c}0}$ measured with the external magnetic field directed along the easy axis of magnetisation. 
The rise of the coercivity observed for an increasing angle $\psi$ followed the Kondorsky 1/cos$\psi$ relation\cite{34,35} indicating that the magnetisation reorientation process takes place by the domain wall motion. 
The small deviation of the coercivity from the 1/cos$\psi$ function seen for larger angles $\psi$ can be related to the presence of domain wall pinning sites.

Atomic and Magnetic Force Microscopy (AFM/MFM) measurements were carried out to obtain more information about domain
structure and domain walls. 
An example of the MFM image obtained for sample irradiated with 100 laser pulses is presented in Fig.~\ref{fig8}a together with a~corresponding AFM image shown in Fig.~\ref{fig8}b.
A cross-section of the line visible in MFM image is presented in Fig.~\ref{fig8}c. 
The asymmetrical shape of the cross-section is the evidence for the presence of magnetic stray fields, which are characteristic for Neel-type domain walls.\cite{39} 
The observed domain wall widths $d_{\mathrm{w}}$, determined from MFM image, were in range from 80~nm to 150~nm with an average value of $125 \pm 31$~nm. 
Using the values of domain wall widths it is possible to estimate a value of the anisotropy energy $K_{\mathrm{u}}$ by applying
the equation:\cite{40,41}
\begin{equation}
 K_{\mathrm{u}} = \pi^{2} \frac{J}{\delta_{\mathrm{w}}^{2}}
\end{equation}
where $J$ is the exchange interaction energy, taken for thin FePd alloy films and equal to $1 \cdot 10^{-11}$~J/m.\cite{40} 
The values of the anisotropy energy $K_{\mathrm{u}}$ for different domain wall widths, calculated from (4), are in the range from $4,4 \cdot 10^{3}$~J/m$^{3}$ to $1,2 \cdot 10^{4}$~J/m$^{4}$, and are about two orders of magnitude lower than anisotropy energy for a~well-ordered L1$_{0}$ FePd alloy.\cite{7,40,42,43} 
Such low anisotropy energy can be a~result of the poor chemical order of the material and the grained sample microstructure with the nanocrystallites separated from each other.

The anisotropy energy could be compared with the magnetostatic energy $E_{\mathrm{m}}$, expressed by the equation
\begin{equation}
 E_{\mathrm{m}} = \frac{1}{2} \mu_{0} M_{\mathrm{s}}^{2} \cos^{2}\psi \ ,
\end{equation}
where $M_{\mathrm{s}}$ is the saturation magnetisation. 
The maximum value of $E_{\mathrm{m}}$ was obtained for the direction perpendicular to the sample plane (angle $\psi=0^{\circ}$). 
Taking into account the measured saturation magnetisation (see Fig.~\ref{fig5}) the magnetostatic energy $E_{\mathrm{m}}$ for angle $\psi=0^{\circ}$ after irradiation with 100 pulses has a value of $7 \cdot 10^6$~J/m$^{3}$. 
The value of the anisotropy energy $K_{\mathrm{u}}$ is larger than the magnetostatic energy $E_{\mathrm{m}}$ for angle $\psi$ greater than 85$^{\circ}$ in agreement with the in-plane easy axis of magnetisation. 
Relatively low anisotropy energy caused the broadening of the domain wall, which also decreased the coercivity field, since the energy cost for domain wall movement was smaller than for well-ordered material with high anisotropy energy.

\vspace{-0.5cm}
\section{Conclusions}

\vspace{-0.2cm}
In this paper we present the results of the research on structural and magnetic ordering induced in Cu/Fe/Pd multilayers by the
pulsed laser annealing. 
The laser annealing allowed us to study the structural and magnetic properties of the material at the early stages of the transformation from a multilayer to an alloy. 
The laser annealing of the Cu/Fe/Pd multilayers led to the formation of the partially L1$_{0}$-ordered FePdCu nanocrystallites. 
It was found, that single laser pulse causes the disappearance of the periodically layered structure of the initial sample, and after application of a~larger number of laser pulses the L1$_{0}$-ordered nanocrystallites started to appear. 
First, as the result of the laser irradiation the (111)-oriented grains raised from the coherent domains present in the initial superlattice, and the annealing process led to partial structural ordering of these crystallites, which was reflected in the decrease of microstrains. 
However, the laser annealing did not change the vertical size of these grains. 
The (002)-oriented grains were formed after at least 10 laser pulses by ordering of the material at the interfaces of the multilayer. 
The further irradiation caused the monotonic increase of the vertical size of these crystallites; however, it did not provide the well-ordered grains. 
The magnetic ordering of the irradiated material was observed as the increase of the saturation magnetisation and the Curie
temperature for increasing number of the laser pulses. 
However, the degree of order induced by laser irradiation was not sufficient to provide the magnetocrystalline anisotropy large enough to overcome the magnetostatic energy, which resulted in orientation of the easy axis of magnetisation in-plane. 
The magnetisation reversal took place by the Neel-type domain wall motion. 
It was also found, that irradiation with 1000 pulses leads to the loss of ordering, which can be attributed to the annealing of the sample at higher temperature than predicted by the theoretical calculations. 
This results in a~partial material ablation.

\noindent
\vspace{0.3cm}

\textbf{Acknowledgements:} This work was partially supported by Polish National Science Center with Contracts No. 2012/07/N/ST8/00533 and 2012/05/B/ST8/01818.

\bibliographystyle{plainnat}
\vspace{1.1ex}
\begin{center}
 $\star$ $\star$ $\star$
\end{center}

\vspace{-9ex}

% styl listowania wpisów bibiograficznych
\setlength{\bibsep}{0pt}
\renewcommand{\bibnumfmt}[1]{$^{#1}$}

\end{document}